# Impact of Brand Dynamics on Insurance Premiums in Turkey

### *By* Yhlas SOVBETOV †

**Abstract.** This paper examines influences of brand dynamics on insurance premium productions in Turkey using a dynamic GMM panel estimation technique sampling 31 insurance firms over 2005-2015. The results reveals that brands trust appears as a chief driving force behind premium production where its unit increase augments premium outputs by 5.32 million Turkish Liras (TL). Moreover, the brand value of firms also appears a statistically significant determinant of premium sales, but its size impact remains limited comparing to brand trust, i.e. a million TL increase in brand value generates only 0.02 million TL increase in sales. On the other hand, the study also documents a strong momentum driven from past years premium production with trade-off magnitude of 1 to 0.85. This might imply a higher loyalty-stickiness of customers in Turkey, as well as a self-feeding *"bandwagon effect"*.

## 1. Introduction

One of the most popular and important marketing concept, that has recognized in 1980s, is a concept of brand value. As an increasing importance of this concept, marketing managers have sought the methods to measure this concept by monitoring its influence and strength. The brand value has been subjected to many researches so far and according to one of them (Wood, 2000), the firm gains a competitive advantage with high brand value. As Knowles (2003) states that this concept cannot be measured with tangible aspects, but it has very strong impact on both awareness and price of a product. When similar competitive products are compared, the increase in price is the brand value. In other words, it is an intangible value added to a product through the brand (Kotler & Keller, 2006).

On the other hand, Berg, Matthews, & O'Hare (2007) define the brand value as, a set of passive and actives such as customer satisfaction, customer loyalty, quality awareness, and uniqueness or distinctiveness of a product. In the same way, Park et al. (2010) defines it as the first thought of the customers about product. Also, they stated that brand value is an intangible value that added to the product which indicates its quality, and meantime this unseen value gives psychological advantage to the customers who bought it. Stahl et al. (2011) defines the brand as an expression of individuals' evaluative judgment to the owner institution. This expression often forms the reputation, prestige, and esteem of the organization, and it helps to retention of organization's current customer portfolio by increasing their loyalty, and to enhance the portfolio by acquisition of new consumers. Therefore, the perception of brand status of firms is valuable marketing tool, however assessing this perception is not too simple as they are intangible assets. Today, brands are no longer just a simple tool of marketing, but they became inevitable basis of marketing. Unlike unbranded products, the branded ones contain an additional economic value which became a core of a number studies in the last decade. For instance, Slotegraaf & Pauwels (2008) investigates impact of brand equity and innovations on sales promotions over more than hundred brands, and find that the brand value has a permanent positive sales effect. More interestingly, they also observe that this effect is greater for brands with higher equity comparing to the brands with low equity. Indeed, this might be addressing Stahl et al.'s (2011) findings about capability of well-established brand value in acquisition of new customers.

† Institute of Social Sciences, Istanbul University, 34452 Beyazıt/Fatih, Istanbul, Turkey.
☎. +90 534 388 69 82
✉. ihlasnobatovich@gmail.com

Well, then how customers perceive the brand equity of firms? Actually, a sizeable investigations (Simon & Sullivan, 1993; Wang et al, 2009; Chi et al., 2009; Sharma et al., 2010; Peterson & Jeong, 2010; Guo et al., 2011) show that individuals judge brands on the basis of firms', primarily, share prices (financially), elegancy (visually), technical superiority (technically), asset size (market dominancy), and mediaticity (awareness/recognition). Indeed, financial side of brand perception is measurable as they are numerically represented. For instance, firm size, capital structure, sales, and dividend payments are some of measurable assets of firms. On the other hand, brand perceptions are formed by more intangible or unmeasurable assets rather than tangible or measurable ones. This fact is documented by Eskildsen et al. (2003) who investigated predictive power of intangible assets, and reported that 22% of market value of firms was comprised of unmeasurable assets in 1978, however it increased to 78% by 2001. Hsu et al. (2013) also reports that the brands form the major part of these unmeasurable assets, and establishing a strong brand would provides competitive advantages and would persuades firms' future earnings through preserving current customer portfolio and effortlessly boosting demand for products.

Mainly two approaches are dominating extant literature of assessment of the brand value: Customer-based Valuation (CBV) and Market-based Valuation (MBV). These approaches are concisely displayed in the figure 1 below. Fayrene & Lee (2011) state that CBV approach measures customers' evaluation judgment about the status of the firm through brand awareness, brand image, and brand loyalty. On the other hand, Christodoulides & de Chernatony (2010) assert that MBV approach computes the firm's brand value through its market determinants such as reputation (dominance), environmental influences, and internal strength. However, Kapferer (2010) argues that CBV would lead to true value only when MBV is accounted too. He tries to combine these two approaches to find the true value by giving more weight to the customer perspective approach. He explains the reason of this as *"the brands have financial value because they have created assets in the minds and hearts of customers"*.

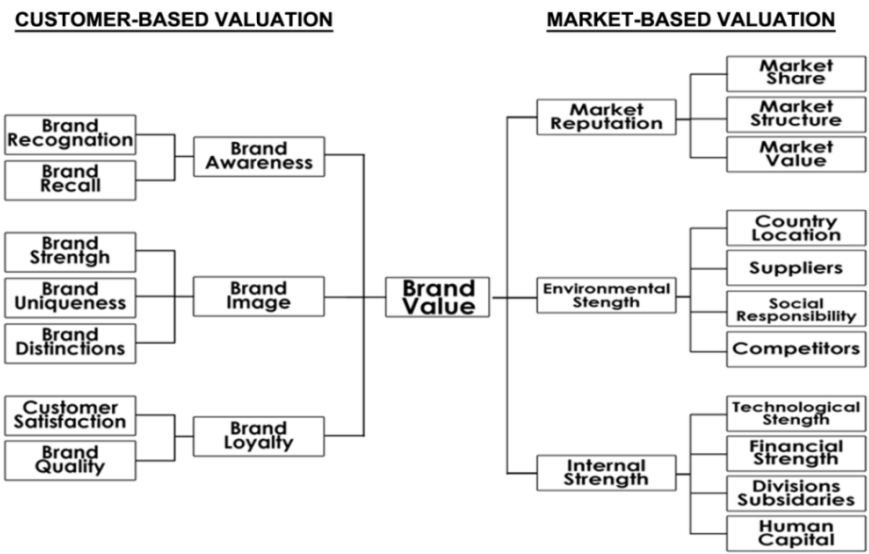

**Figure 1.** *Customer- and Firm-Based Brand Valuation of Brand Value*
**Source:** Chieng Fayrene & Lee (2011) and Christodoulides & de Chernatony (2010)

Today firms are seeking ways of creating intangibles assets such as brand value rather than tangible ones to increase their competitive power over the rest market participants. Referring to Kapferer (2007), the brand is a powerful marketing factor that creates the assets in the thoughts, minds, and hearts of the customers. Physically, brand is just a name, whereas, financially it is more than the logo or writings on it (Kayali et al. 2004).

A number of studies have been pursued to find out impact of brand influences on other financial indicators. For instance, Chang & Wildt (1994) find out that the perceived value of product by customers is positively related with their willingness of purchase, and so sales. Equally, Pappu et al. (2005) states that the main determinant of brand value is the customer itself, because a new customer most likely to prefer a firm that already has large customer portfolio. In other words, if number of existing customers a firm is large than others, then new customers generally perceive that the firm has the most valuable brand in that market. On other hand, Grace & O'Cass (2004) argue that the value of a brand will be materialized from customers experiences in a long-term. Latterly findings of Poppu et al. (2005) are confirmed by Bilgili et al. (2008) who investigate impact of brand value in Turkish insurance industry, and reports that all insured individuals in one family are often customers of the same company. Moreover, they observe that customers in Turkey are too loyal that their willingness to shift the company is too less. They state that the main determinant of customers retention is favorable prices, not the brand value. Because according their survey, 55% of customers haven't agreed to pay an extra money for the brand of their current company comparing to other firms' brands. Employing correlation analysis, they find a strong positive relationship between brand value and perceived quality, and weak positive relationships between brand value and brand recognition, brand recall, and brand loyalty.

On the other hand, Roy (2012) in his research about brand loyalty in Indian insurance companies aims to measure the perceived quality which was set forth as a main determinant of brand value. He observes that nearly 79% of customers think the service which they get meets their expectations in terms of money they pay. Nearly 52% of the same group also reports that the attitude of insurance firms towards the problems is positive and very efficient. However, when Roy asks if they would be loyal to their current insurance firms or not, interestingly just 37% agrees to be loyal in long term, while 34% disagrees, and 29% are not sure about it. All in all, he concludes that brand value is majorly determined by perceived quality, and the role of brand loyalty is limited.

## 2. Outlook of Turkish Insurance Industry

The data reports IAT[2] reveals that the top 10 insurance firms have produced 21.69 billion Turkish Liras (TL) in 2015 (table 1) which represents 69.89% of total industry production. More specifically, Allianz Sigorta has been leading Turkish insurance industry since 2013 -after the date when they bought out Yapi Kredi Sigorta which was 5th largest in the industry- ceasing dominance of Anadolu sigorta. Today, Allianz Sigorta owns 14.85% of total Turkish insurance market where Anadolu, Axa, Mapfre, and Aksigorta follow with 12.93%, 9.95%, 6.84%, and 5.23% market shares respectively. More importantly, table 3 shows that Mapfre Genel Sigorta, Halk Sigorta, and Sompo Japan Sigorta have been growing aggressively for last few years whereas Axa, Ak sigorta, and Ziraat Sigorta have been experiencing negative growths.

**Table 1.** *Premium Production of Turkish Insurance Firms during 2005-2015 period (TL millions)*

| | COMPANY NAME | 2005 | 2006 | 2007 | 2008 | 2009 | 2010 | 2011 | 2012 | 2013 | 2014 | 2015 |
|---|---|---|---|---|---|---|---|---|---|---|---|---|
| 1 | Ace European Group Ltd. | - | - | - | - | 3.78 | 22.23 | 24.43 | 36.92 | 56.50 | 60.12 | 94.72 |
| 2 | Acıbadem Sağlık ve Hayat Sigorta | 65.89 | 64.39 | 91.30 | 107.99 | 127.00 | 135.20 | 176.82 | 237.22 | 308.02 | 416.31 | 522.36 |
| 3 | Aegon Emeklilik ve Hayat | 13.03 | 10.66 | 9.30 | 7.48 | 7.26 | 14.35 | 29.93 | 55.55 | 89.23 | 133.86 | 195.62 |
| 4 | Aig (American Life) Hayat | 62.92 | 83.17 | 110.14 | 101.28 | - | - | - | - | - | - | - |
| 5 | AIG Sigorta | 94.84 | 121.91 | 154.40 | 186.37 | - | - | 199.69 | 217.52 | 275.78 | 300.30 | 342.28 |
| 6 | Ak Emeklilik | 54.81 | 61.32 | - | - | - | - | - | - | - | - | - |
| 7 | Aksigorta | 516.42 | 651.44 | 793.57 | 829.21 | 851.17 | 886.29 | 1,136.74 | 1,311.33 | 1,526.14 | 1,713.62 | 1,622.19 |
| 8 | Allianz Hayat ve Emeklilik | 89.27 | 87.90 | 82.46 | 78.60 | 77.75 | 74.85 | 84.03 | 91.04 | 99.89 | 109.55 | 115.50 |
| 9 | Allianz Sigorta | 636.34 | 765.04 | 860.81 | 905.22 | 930.69 | 995.78 | 1,129.04 | 1,444.88 | 3,199.13 | 3,216.18 | 4,050.87 |
| 10 | Allianz Yaşam ve Emeklilik | - | - | - | - | - | - | - | - | 244.06 | 337.09 | 440.71 |
| 11 | Anadolu Sigorta | 825.93 | 1,030.37 | 1,192.59 | 1,161.39 | 1,243.48 | 1,420.46 | 1,926.09 | 2,234.63 | 2,749.74 | 3,004.83 | 3,610.67 |
| 12 | Anadolu Hayat Emeklilik | 348.48 | 340.79 | 339.19 | 345.33 | 500.05 | 357.61 | 348.44 | 367.97 | 395.01 | 365.68 | 402.55 |
| 13 | Ankara Sigorta | 172.42 | 180.81 | 192.12 | 183.74 | 196.09 | 215.78 | 141.24 | 152.52 | 173.99 | 221.53 | 193.01 |
| 14 | Asya Emeklilik ve Hayat | - | - | - | - | - | - | - | - | 0.48 | 2.64 | 5.15 |
| 15 | Atradius Credit Insurance NV | - | - | 0.92 | 3.75 | 3.78 | 5.15 | 8.07 | 9.19 | 11.94 | 16.59 | 31.51 |
| 16 | Aviva Sigorta | 142.86 | 180.43 | 223.22 | 251.62 | 274.78 | 281.81 | 300.94 | 340.06 | 271.15 | 190.51 | - |
| 17 | AvivaSA Emeklilik ve Hayat | 157.06 | 131.47 | 79.92 | 150.51 | 154.92 | 155.31 | 148.42 | 197.54 | 232.89 | 258.31 | 263.46 |
| 18 | Axa (Oyak) Hayat ve Emeklilik | 113.26 | 127.22 | 140.85 | 123.61 | 108.62 | 56.65 | 54.84 | 62.63 | 99.02 | 36.24 | 23.02 |
| 19 | Axa (Oyak) Sigorta | 759.85 | 917.49 | 1,129.74 | 1,234.02 | 1,277.19 | 1,518.55 | 1,997.61 | 2,386.25 | 3,168.37 | 3,078.12 | 3,065.69 |
| 20 | Batı | -0.47 | 0.00 | 0.01 | 0.00 | - | - | - | - | - | - | - |
| 21 | BNP Paribas Cardif Emeklilik | - | - | - | - | - | - | 10.88 | 53.57 | 75.14 | 134.55 | 145.72 |
| 22 | BNP Paribas Cardif Hayat | - | - | 0.00 | 11.07 | 44.61 | 36.75 | 77.80 | 40.88 | 58.20 | 64.94 | 86.00 |
| 23 | BNP Paribas Cardif Sigorta | - | - | 0.00 | 0.30 | 11.51 | 15.99 | 11.27 | 11.14 | 22.26 | 27.57 | 61.51 |
| 24 | Chartis Sigorta | - | - | - | - | 174.45 | 173.59 | - | - | - | - | - |
| 25 | Cigna Finans Emeklilik ve Hayat | 161.30 | 206.57 | 0.97 | 47.90 | 48.78 | 85.49 | 124.68 | 161.12 | 202.00 | 209.95 | 254.14 |
| 26 | Cigna Hayat Sigorta | - | - | - | - | - | - | 0.22 | 0.27 | -0.02 | 0.00 | - |
| 27 | CIV Hayat Sigorta | - | - | 0.00 | 5.64 | 13.69 | 25.63 | 29.47 | 38.81 | 12.79 | -0.19 | 0.00 |
| | COMPANY NAME | 2005 | 2006 | 2007 | 2008 | 2009 | 2010 | 2011 | 2012 | 2013 | 2014 | 2015 |
| 28 | Coface Sigorta | - | - | 7.11 | 16.20 | 11.98 | 17.87 | 23.80 | 32.21 | 48.92 | 59.32 | 71.16 |
| 29 | Demir Hayat Sigorta | 0.00 | 0.00 | 0.00 | 0.52 | 3.71 | 10.47 | 17.14 | 19.51 | 50.99 | 55.21 | 58.90 |
| 30 | Demir Sigorta | 20.85 | 28.84 | 32.87 | 37.28 | 41.95 | 52.00 | 50.25 | 59.02 | 28.73 | 31.28 | 41.05 |
| 31 | Deniz Hayat (Global Hayat) | 2.55 | 3.24 | 23.08 | 43.38 | 55.86 | 72.44 | - | - | - | - | - |
| 32 | Dubai Starr Sigorta | - | - | 0.00 | 0.61 | 75.81 | 128.02 | 115.65 | 70.77 | 80.03 | 124.65 | 162.96 |
| 33 | Ege (Euro) Sigorta | 4.88 | 4.46 | 4.37 | 12.84 | 57.64 | 58.89 | 100.33 | 176.85 | 205.31 | 246.47 | 198.76 |
| 34 | Ergo (İsviçre) Emeklilik ve Hayat | 62.44 | 64.67 | 61.43 | 42.68 | 23.34 | 26.06 | 24.03 | 21.99 | 13.27 | 8.50 | 7.55 |
| 35 | Ergo (İsviçre) Sigorta | 418.32 | 513.83 | 636.65 | 698.30 | 675.73 | 693.65 | 699.97 | 683.28 | 564.05 | 673.09 | 902.82 |
| 36 | Euler Hermes Sigorta | - | - | - | - | - | - | - | 11.75 | 25.16 | 35.38 | 65.02 |
| 37 | Eureko Sigorta AŞ (Garanti) | 298.44 | 357.79 | 415.47 | 478.52 | 539.18 | 618.40 | 709.04 | 685.40 | 789.08 | 801.15 | 1,002.55 |
| 38 | Fiba Emeklilik ve Hayat | - | - | - | - | - | - | - | - | 7.42 | 9.02 | 15.93 |
| 39 | Fortis Emeklilik ve Hayat | 17.62 | 25.86 | 19.37 | 28.65 | 23.66 | 19.62 | - | - | - | - | - |
| 40 | Garanti Emeklilik ve Hayat | 55.85 | 98.60 | 108.05 | 123.54 | 181.10 | 234.16 | 240.49 | 262.86 | 298.11 | 318.76 | 328.80 |
| 41 | Genel Yaşam | 64.67 | 82.24 | 111.08 | 125.69 | - | - | - | - | - | - | - |
| 42 | Generali Sigorta | 59.81 | 69.13 | 72.13 | 97.19 | 84.74 | 85.24 | 106.78 | 83.82 | 87.69 | 158.03 | 199.79 |
| 43 | Groupama Emeklilik (Başak) | 111.08 | 184.35 | 215.68 | 295.64 | 363.52 | 51.62 | 75.90 | 93.81 | 110.40 | 123.45 | 126.58 |
| 44 | Groupama Sigorta (Başak) | 374.68 | 434.09 | 467.99 | 524.46 | 726.68 | 693.87 | 818.26 | 826.80 | 975.76 | 1,057.75 | 1,114.56 |
| 45 | Güneş Sigorta | 446.80 | 495.44 | 638.14 | 709.62 | 727.07 | 737.37 | 819.95 | 922.46 | 1,076.72 | 1,212.63 | 1,288.39 |

[2] Insurance Association of Turkey (IAT). (a) Balance Sheet of Insurance Reinsurance Companies: [Retrieved from].
(b) Financial Profit Loss Account of Insurance Reinsurance Companies: [Retrieved from].
(c) Technical Profit Loss Account of Insurance Reinsurance Companies: [Retrieved from].
(d) Official Statistics: [Retrieved from].

| | | 2005 | 2006 | 2007 | 2008 | 2009 | 2010 | 2011 | 2012 | 2013 | 2014 | 2015 |
|---|---|---|---|---|---|---|---|---|---|---|---|---|
| 46 | Güven | 167.87 | 188.45 | 222.89 | 221.07 | - | - | - | - | - | - | - |
| 47 | Güven Hayat | 17.41 | 22.02 | 22.76 | 29.31 | - | - | - | - | - | - | - |
| 48 | Halk Hayat ve Emeklilik | 13.78 | 33.52 | 46.72 | 73.85 | 77.89 | 141.34 | 183.96 | 181.46 | 281.51 | 238.35 | 317.75 |
| 49 | Halk Sigorta | 66.59 | 108.07 | 113.62 | 105.71 | 113.35 | 155.69 | 206.50 | 395.16 | 470.36 | 537.82 | 757.37 |
| 50 | HDI Sigorta | 65.35 | 113.02 | 157.75 | 151.26 | 175.30 | 223.53 | 287.93 | 398.55 | 476.28 | 584.41 | 700.68 |
| 51 | Hür Sigorta | 34.28 | 46.50 | 42.05 | 46.24 | 52.87 | 52.80 | 50.09 | 57.02 | 43.93 | 20.90 | 2.29 |
| 52 | NN Hayat ve Emeklilik (ING) | - | - | - | - | 0.00 | 20.32 | 40.13 | 53.50 | 68.97 | 76.31 | 97.51 |
| 53 | Inter | 0.00 | 0.00 | 0.00 | 0.00 | - | - | - | - | - | - | - |
| 54 | Işık Sigorta | 61.85 | 81.97 | 105.49 | 109.36 | 112.35 | 118.24 | 138.05 | 165.66 | 184.07 | 164.44 | 157.93 |
| 55 | Katılım Emeklilik ve Hayat | - | - | - | - | - | - | - | - | 0.00 | 3.39 | 18.20 |
| 56 | Liberty Sigorta | - | - | 152.12 | 77.98 | 46.35 | 59.78 | 93.46 | 161.38 | 150.66 | 135.16 | 158.77 |
|  | COMPANY NAME | 2005 | 2006 | 2007 | 2008 | 2009 | 2010 | 2011 | 2012 | 2013 | 2014 | 2015 |
| 57 | Mapfre Genel Sigorta | - | - | - | - | - | - | 557.79 | 886.59 | 1,352.89 | 1,490.95 | 2,110.57 |
| 58 | Mapfre Genel Yaşam Sigorta | 206.77 | 263.30 | 321.76 | 345.68 | 361.16 | 411.79 | 128.26 | 10.74 | 11.43 | 12.75 | 13.74 |
| 59 | Merkez | 0.00 | 0.00 | 0.00 | 0.00 | - | - | - | - | - | - | - |
| 60 | Metlife Emeklilik ve Hayat | - | - | - | - | 131.79 | 157.90 | 192.94 | 170.99 | 233.26 | 256.54 | 321.09 |
| 61 | Neova Sigorta | - | - | - | - | 0.26 | 61.80 | 84.93 | 131.14 | 256.80 | 380.33 | 514.43 |
| 62 | New Life Yaşam Sigorta | 0.00 | 0.00 | 3.40 | 5.84 | 6.02 | 3.36 | 0.82 | 0.47 | 0.51 | 0.33 | 0.20 |
| 63 | Orient Sigorta | - | - | - | - | - | - | - | - | - | 8.10 | 60.51 |
| 64 | Ray Sigorta | 200.75 | 254.27 | 270.99 | 274.20 | 253.83 | 252.37 | 254.41 | 302.95 | 354.30 | 380.03 | 444.64 |
| 65 | Rumeli | 0.00 | 0.00 | 0.00 | 0.00 | - | - | - | - | - | - | - |
| 66 | Rumeli Hayat Sigorta | 1.26 | 1.07 | 0.74 | 0.40 | 0.27 | 0.20 | 0.12 | 0.06 | 0.02 | 0.02 | 0.01 |
| 67 | SBN Sigorta (Şeker) | 87.22 | 131.37 | 3.63 | 58.32 | 84.95 | 109.93 | 68.71 | 60.99 | 83.94 | 100.52 | 125.26 |
| 68 | Sompo Japan Sigorta (Fiba) | - | - | 289.33 | 313.71 | 305.72 | 318.58 | 329.70 | 450.77 | 555.81 | 686.79 | 1,063.47 |
| 69 | SS Doğa Sigorta Kooperatifi | - | - | - | - | - | - | - | - | 0.00 | 26.48 | 291.33 |
| 70 | SS Koru Sigorta Kooperatifi | - | - | - | - | - | - | 2.09 | 29.66 | 33.93 | 29.51 | 95.26 |
| 71 | Teb | 73.32 | 113.91 | - | - | - | - | - | - | - | - | - |
| 72 | Ticaret | 6.07 | 0.00 | - | - | - | - | - | - | - | - | - |
| 73 | Turins Sigorta | - | - | - | - | - | - | - | - | 0.90 | 4.24 | 4.12 |
| 74 | Türk Nippon Sigorta | 0.03 | 0.00 | - | - | 4.19 | 21.11 | 35.93 | 30.39 | 44.11 | 71.89 | 110.34 |
| 75 | Türk P&I Sigorta | - | - | - | - | - | - | - | - | 0.00 | 7.20 | 13.87 |
| 76 | Vakıf Emeklilik | 74.97 | 86.10 | 81.63 | 63.45 | 68.87 | 92.99 | 141.27 | 173.25 | 225.60 | 185.54 | 243.65 |
| 77 | Unico Sigorta | - | - | - | - | - | - | - | - | - | - | 221.83 |
| 78 | Yapı Kredi | 469.74 | 574.34 | 628.14 | 631.54 | 607.98 | 758.17 | 973.10 | 1,227.38 | - | - | - |
| 79 | Yapı Kredi Emeklilik | 117.04 | 112.69 | 98.68 | 110.23 | 92.38 | 109.96 | 176.00 | 212.48 | - | - | - |
| 80 | Ziraat Hayat ve Emeklilik | - | - | - | - | 0.00 | 601.77 | 810.09 | 590.37 | 804.82 | 617.49 | 630.50 |
| 81 | Ziraat Sigorta | - | - | - | - | 0.00 | 172.59 | 318.46 | 378.78 | 567.63 | 703.21 | 932.35 |
|  | **TOTAL SECTOR** | **7,816.49** | **9,454.10** | **10,931.47** | **11,774.22** | **12,436.06** | **14,129.39** | **17,165.08** | **19,826.65** | **24,229.62** | **25,989.55** | **31,025.90** |

On the other hand, Brand-Finance consulting firm estimates that Allianz Sigorta is the most valued brand in Turkish insurance market with 48.89 billion TL which is followed by Axa Sigorta with 45.61 billion TL, Anadolu Sigorta with 38.19 billion TL, Ak Sigorta with 23.40 billion TL, and Groupama Sigorta with 15.49 billion TL. Moreover, Ak Sigorta appears as the best performer in terms of brand value growth among top 10 premium producers in Turkey, being increased its brand value almost 2 fold (1.87 times) since 2005. Although, Ak Sigorta fails to perform same performance in term of generating premiums, the figure 2 shows a tight correlation between premium production and brand values of top 10 premium producer insurance firms.

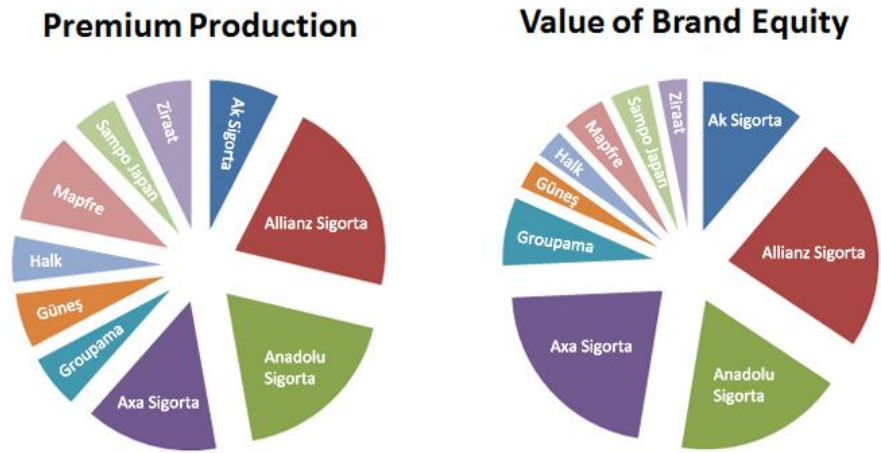

**Figure 2.** *Premium Production and Value of Brand Equity of Top 10 Turkish Insurance Firms in 2015*
**Source:** IAT & Brand-Finance.

## 3. Data & Methodology

The literature about impact of brand dynamics on sales is sizeable enough. However, majority of the literature omits insurance industry rather focuses on food and beverages, apparel and accessories, manufacturing, electronics, automobile, and other consumer cyclicals and non- cyclicals. This paper aims to contribute to the neglected part of the literature by empirically analyzing influences of brand dynamics on insurance premium productions over 2005-2015 by sampling 31 firms that operate in Turkey. For analysis, I consider a dynamic panel model with Generalized Method of Moments (GMM) approach following Holtz-Eakin, Newey, & Rosen (1988) and Arellano & Bond (1991).

$$PP_{a,t} = \beta_0 + \sum_{i=1}^{n} \beta_{a,i} PP_{a,t-i} + \sum_{i=0}^{m} \beta_{a,n+1+i} BV_{a,t-i} + \sum_{i=0}^{k} \beta_{a,n+m+2+i} BT_{a,t-i} + \varepsilon_{a,t} \qquad (1)$$

where $\beta_0$ is an intercept; $PP_{a,t}$ is a premium production of insurance firm "a" at time "t"; $BV$ and $BT$ are brand value and brand trust rating of the firm; and "$\varepsilon$" is a residual term of the model. The data for $PP$ is gathered from statistic report of Insurance Association of Turkey [3](IAT), while data for $BV$ and $BT$ are derived from annual reports of Brand-Finance [4]institution. The Brand-Finance (BF) is a London-based independent brand valuation and strategy consultancy that acts over 20 countries worldwide, and estimates brand-related data of country- and industry- specific companies using the Royalty Relief methodology. This approach calculates value of a firm which would be willing to pay for licensing its brand assuming that the firm do not own it. In the first step, BF determines strength of a brand using a balanced scorecard focusing on firm's financial performance and sustainability. Secondly, it determines appropriate royalty rate using their own extensive database of license agreements. In the final step, BF forecast brand specific revenues to derive brand values, and discounts it to present day as net present value (NPV), which will also equate to the firm's brand value.

Moreover, in the model (1) I restrict the "t" with 2005-2015 periods, and "a" with 31 insurance firms. To determine maximum size sample for this study, I scanned whole Turkish insurance markets, and found out 83 different insurance firms those have/have been operated/operating since 2005 up to date. However, my final sample size is reduced to 31 firms due to data limitations (table 2).

---

[3] Insurance Association of Turkey (IAT). (a) Balance Sheet of Insurance Reinsurance Companies: [Retrieved from].
(b) Financial Profit Loss Account of Insurance Reinsurance Companies: [Retrieved from].
(c) Technical Profit Loss Account of Insurance Reinsurance Companies: [Retrieved from].
(d) Official Statistics: [Retrieved from].
[4] Brand-Finance. (a) Most Valuable Brands in Insurance: [Retrieved from].
(b) Best Brands in Turkey: [Retrieved from].

**Table 2.** *Sample size of the study*

| | COMPANY NAME | | COMPANY NAME |
|---|---|---|---|
| 1 | Acıbadem Sağlık ve Hayat Sigorta AŞ | 17 | Garanti Emeklilik ve Hayat AŞ |
| 2 | Aegon Emeklilik ve Hayat AŞ | 18 | Generali Sigorta AŞ |
| 3 | AIG Sigorta AŞ | 19 | Groupama Sigorta AŞ (Başak) |
| 4 | Aksigorta AŞ | 20 | Gunes Sigorta AŞ |
| 5 | Allianz Sigorta AŞ (Koç) | 21 | Halk Sigorta AŞ |
| 6 | Anadolu Anonim Türk Sigorta Şirketi | 22 | HDI Sigorta AŞ |
| 7 | Ankara Sigorta AŞ | 23 | Işık Sigorta AŞ |
| 8 | Aviva Sigorta AŞ | 24 | Mapfre Genel Sigorta AŞ |
| 9 | Axa (Oyak) Sigorta AŞ | 25 | Ray Sigorta AŞ |
| 10 | BNP Paribas Cardif Sigorta AŞ | 26 | SBN (Şeker) Sigorta AŞ |
| 11 | Cigna Hayat Sigorta AŞ | 27 | Sompo Japan Sigorta AŞ (Fiba) |
| 12 | Demir Sigorta AŞ | 28 | Yapi Kredi Sigorta AŞ |
| 13 | Dubai Starr Sigorta AŞ | 29 | Vakıf Emeklilik Sigorta AŞ |
| 14 | Ege (Euro) Sigorta AŞ | 30 | Ziraat Sigorta AŞ |
| 15 | Ergo (İsviçre) Sigorta AŞ | 31 | Zurich Sigorta AŞ |
| 16 | Eureko Sigorta AŞ (Garanti) | | |

To summarize the data, I schedule a descriptive analysis in table 3 regarding abovementioned data series where *PP* and *BV* are given in million TL, and *BT* is given in numerical scale that is transformed from letter rating grades by using a methodology explained in Appendix-A.

**Table 3.** *Descriptive Statistics of Data*

| | PP | BV | BT |
|---|---|---|---|
| Mean | 548.45 | 7,496.13 | 79.12 |
| Median | 281.71 | 4,976.98 | 82.5 |
| Maximum | 4,607.09 | 48,898.36 | 97.5 |
| Minimum | 0.97 | 694.11 | 47.5 |
| Standard Deviation | 698.58 | 7,666.26 | 10.6 |
| Skewness | 2.73 | 2.63 | -0.69 |
| Kurtosis | 12.09 | 11.09 | 2.95 |
| Observation | 320 | 320 | 320 |

**Notes:** PP and BV are presented in million TL. However, BT is transformed into numerical scale from letter rating grades by using methodology explained in Appendix-A.

Coming to the model (1), the literature of panel model urges two type of effects specification: Fixed effects (FE) and Random effects (RE). In order to find out the most appropriate dynamic panel model for my sample, I shall look to consistency and efficiency of GMM estimators through cross-section FE and RE specifications where both have potential advantages/disadvantages. FE model assumes heterogeneity among constituents of the sample by allowing to have their own intercept values. Although, this intercept differs among entities, it does not change over the time. Therefore, FE model generates unbiased estimates of $β_i$, but it may suffer from high variance due to a larger variation between sample firms. In this case, our dynamic panel model with FE specification becomes as below.

$$PP_{a,t} = β_0 + \sum_{a=1}^{31} α_a D_a + \sum_{i=1}^{n} β_{a,i} PP_{a,t-i} + \sum_{i=0}^{m} β_{a,n+1+i} BV_{a,t-i} + \sum_{i=0}^{k} β_{a,n+m+2+i} BT_{a,t-i} + ε_{a,t} \quad (2)$$

where $D_a$ is a dummy variable which equates 1 for the firm *"a"*, and zero for others in the sample. I also could include a fixed effect for period by considering a dummy variable for years as *"$D_t$"* only in case when the period is different for countries in the sample. But it is not a case for this study.

On the other hand, RE model heals the high variance problem by generating estimates closer, on average, to the true value of any particular country as below.

$$PP_{a,t} = \beta_{a,0} + \sum_{i=1}^{n} \beta_{a,i} PP_{a,t-i} + \sum_{i=0}^{m} \beta_{a,n+1+i} BV_{a,t-i} + \sum_{i=0}^{k} \beta_{a,n+m+2+i} BT_{a,t-i} + \varepsilon_{a,t} \quad (3)$$

$$\beta_{a,0} = \beta_0 + \omega_a \qquad \text{where} \quad \omega_a \sim N(0, \sigma^2)$$

When $\beta_{a,0}$ is plugged into (1) equation model, it becomes as below.

$$PP_{a,t} = \beta_0 + \sum_{i=1}^{n} \beta_{a,i} PP_{a,t-i} + \sum_{i=0}^{m} \beta_{a,n+1+i} BV_{a,t-i} + \sum_{i=0}^{k} \beta_{a,n+m+2+i} BT_{a,t-i} + \omega_a + \varepsilon_{a,t}$$

$$PP_{a,t} = \beta_0 + \sum_{i=1}^{n} \beta_{a,i} PP_{a,t-i} + \sum_{i=0}^{m} \beta_{a,n+1+i} BV_{a,t-i} + \sum_{i=0}^{k} \beta_{a,n+m+2+i} BT_{a,t-i} + u_{a,t} \quad (4)$$

where $u_{a,t} = \omega_a + \varepsilon_{a,t}$. Nevertheless, the estimates of this model is often biased due to potential correlation between covariates of explanatory variables ($\beta_i$) and $\omega_a$. Unlike FE model, it captures both *"within"* and *"between"* deviations, and allows all entities to have a common mean value for intercept. With other words, the dummy variable *"$D_a$"* -was a part of intercept in the FE- becomes a part of error *"$\varepsilon_a$"* in the RE model.

The trade-off between biasedness of RE and high variance problem of FE models is entirely based on researchers' preference and decision. On the other hand, I also can employ *"difference"* or *"orthogonal deviation"* transformation methodology to the specification of my dynamic panel model to remove fixed effects. But Hayakawa (2009) shows that GMM estimator of the model transformed by the forward orthogonal deviation (OD) tends to work better than that transformed by the first difference (FD). Moreover, Arellano & Honore (2001) state that the forward orthogonal deviation method does not only eliminate fixed effect by taking first differences, but also removes serial correlation induced by differencing. So, following Hayakawa (2009), I apply orthogonal deviation transformation, and my (2) model becomes as (5).

$$\widehat{PP}_{a,t} = \sum_{i=1}^{n} \beta_{a,i} \widehat{PP}_{a,t-i} + \sum_{i=0}^{m} \beta_{a,n+1+i} \widehat{BV}_{a,t-i} + \sum_{i=0}^{k} \beta_{a,n+m+2+i} \widehat{BT}_{a,t-i} + \varepsilon_{a,t} \quad (5)$$

where *"hat"* denotes orthogonal deviation transformation.

## 4. Analysis

Although I select OD (in Eq.5) as most appropriate model for this studies, I also present results of Pooled, RE (in Eq.2), FE (in Eq.3), and FD for comparative discussion in table 4. Prior to analysis, I determined the lag length considering information criterion tests such as Akaike, Schwarz, and Hannan-Quinn where all of them urge that the lag length for autoregressive term is 1, and for other (*BV* and *BT*) are zero.

**Table 4.** *Results of Dynamic Panel GMM Analysis*

| Variable | Cross-section Specification | | | | |
|---|---|---|---|---|---|
| | Pooled | FE | RE | OD | FD |
| PP(-1) | 0.8788*** | 1.0530*** | 0.8788*** | 0.8485*** | 0.8515*** |
| | (0.1539) | (0.4156) | (0.1539) | (0.0029) | (0.0055) |
| | [5.7101] | [2.5334] | [5.7101] | [290.6894] | [155.8356] |
| BV | 0.0206** | 0.0084 | 0.0206** | 0.0208*** | 0.0208*** |
| | (0.0101) | (0.0265) | (0.0101) | (0.0000) | (0.0002) |
| | [2.0371] | [0.3176] | [2.0371] | [366.6883] | [103.3887] |
| BT | 0.9644 | 3.3081*** | 0.9644 | 5.3154*** | 5.3497*** |
| | (0.8408) | (0.8740) | (0.8408) | (0.1443) | (0.1846) |
| | [1.1469] | [3.7849] | [1.1469] | [36.8247] | [28.9859] |
| intercept | -96.1434* | -291.2198*** | -96.1434* | - | - |
| | (58.5515) | (68.8748) | (58.5515) | | |
| | [-1.6420] | [-4.2282] | [-1.6420] | | |
| R-square | 0.9469 | 0.9603 | 0.9469' 0.9469" | 0.9873 | 0.9710 |
| J-statistics | 4.2184 | 2.3039 | 4.2184 | 29.2175 | 32.7110 |
| | (0.2388) | (0.5118) | (0.2388) | (0.4538) | (0.2895) |

**Notes:** Each column presents results for different effect-specified panel GMM analysis. All panels include 258 unbalanced observations structured by 31 cross-sections and 11 periods starting from 2005 up to 2015. Pooled panel does not have any effect specifications, whereas FE and RE have cross-sectional fixed and random effect specifications respectively considering static instruments of 2 lags of BV and BT as well as intercept. OD and FD models take into account orthogonal deviations and first-differences of cross-sections respectively considering Arellano-Bond type dynamic instruments of @dyn(pp,-2) @dyn(bv,-2) @dyn(bt,-2), i.e. all instruments have 2 lags. None of the panel models include period effect specifications. Coefficient standard errors and covariance have White characteristics with Arellano-Bond n-step GMM weighting matrix presented in parenthesis. As well as, computed t-statistics for each estimation are displayed in brackets. The diagnostics of each analysis is presented in the bottom part of the table where J-statistics are given with probability of rejection of the null hypothesis (overidentification) in the parenthesis. Two r-square values are shown in RE model where a single and double quotation indicate weighted and unweighted R-square values respectively.

Moreover, to overcome heteroskedasticity problem in residuals, I assign White-robust standard errors and covariances. Additionally, I schedule Arellano-Bond serial correlations test for OD and FD model, and the test indicates that the consecutive residuals are not serially correlated with each other as well as included regressors, thus I concur that both models is purely healthy.

Coming to the results, notice that both weighted and unweighted R-square values of RE is exactly same. This occurs when entire variance of RE is comprised by idiosyncratic effects leaving zero shares for cross-sectional effects. In other words, $\sigma_u$ (cross-section random) in Swamy-Arora estimators of variance components of RE gets rho number of zero, while $\sigma_e$ (idiosyncratic random) gets rho number of 1, i.e. or 100%. In such a case, notice that RE generates exactly same estimation coefficients with the pooled one, as well as application of Hausman test to select RE or FE becomes invalid.

Thus, following Arellano & Honore (2001) I re-estimate panel model applying OD and FD transformations, and I derive more plausible results. Both models generate quite similar estimations, however, as Hayakawa (2009) states OD tends to work better than FD. Notice that both models reveal that autoregressive term and brand-related factors are statistically significant at 1% level. However, t-statistics of estimates are remarkably higher in OD where autoregressive term and brand-related factors with Arellano-Bond type dynamic panel instruments of *PP*, *BV*, and *BT* with 2 lags account 98.73% of variations in dependent variable *PP*. In case of FD, the same regressors account only 97.10% of variation in *PP*. The fitted values of these two models are portrayed in figures 3 and 4 below, which are generated by Eviews 9 software. Additionally, I re-transform these OD and FD transformed actual and fitted values back into their raw initial units, and portray their goodness-of-fit in figure 5.

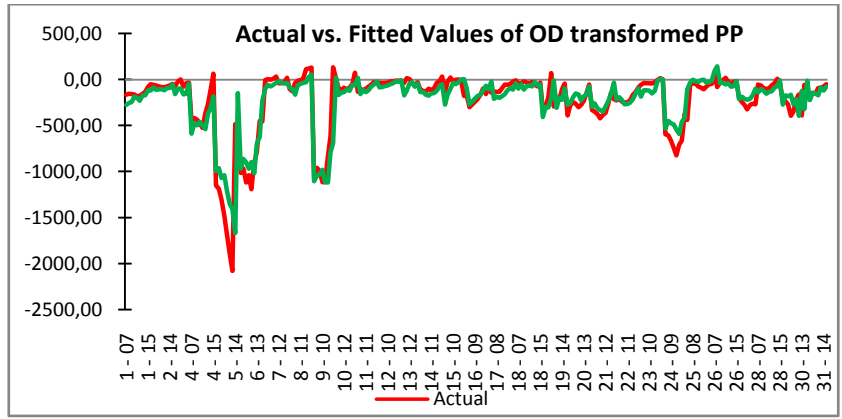

**Figure 3.** *Actual and Fitted Values of Orthogonal Deviation (OD) transformed PP*
**Notes:** The panel includes 258 unbalanced observations structured by 31 cross-sections and 11 periods from 2005 to 2015.

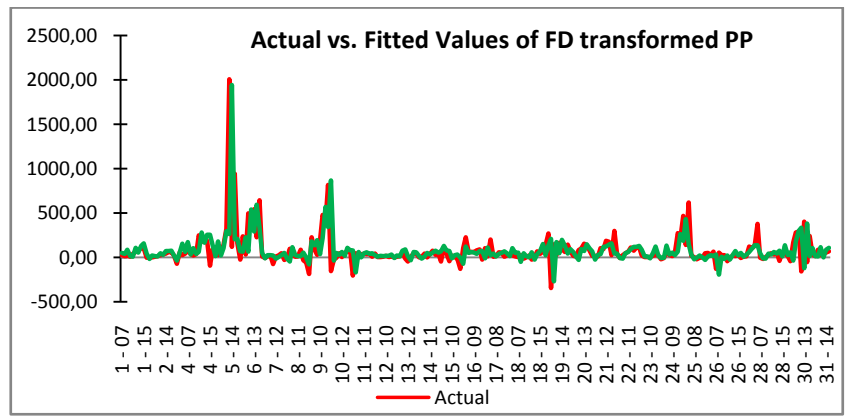

**Figure 4.** *Actual and Fitted Values of First Differenced (FD) transformed PP*
**Notes:** The panel includes 258 unbalanced observations structured by 31 cross-sections and 11 periods from 2005 to 2015.

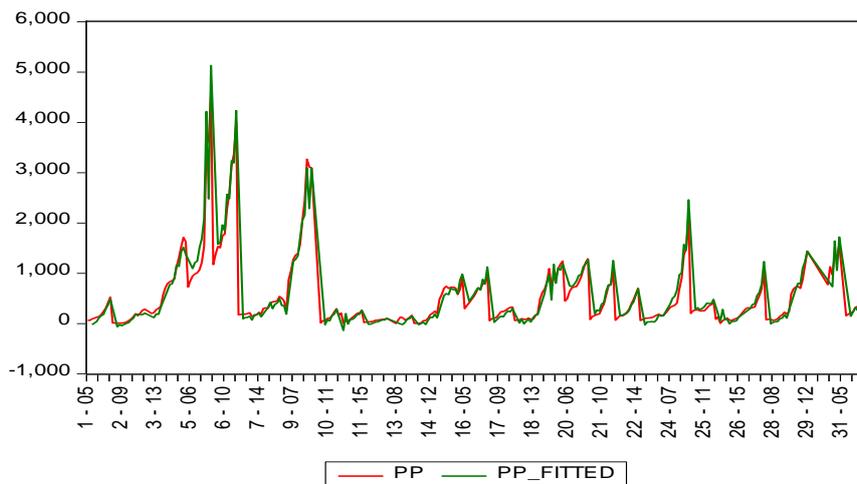

**Figure 5.** *Actual and Fitted Values of PP without any transformation*
**Notes:** The panel includes 258 unbalanced observations structured by 31 cross-sections and 11 periods from 2005 to 2015.

More importantly, null-rejection probability of J-statistics indicate that both OD and FD are feasible, and do not comprise overidentification problem caused by redundant instruments. In turn, it also confirms validity of all used Arellano-Bond type dynamic instruments that help to overcome endogenity problem caused by autoregressive term in the model.

Coming to coefficients, I believe that the positive sign estimate of autoregressive term implies a strong momentum with 0.85 magnitude, and estimated coefficient of *BV* indicates the impact running from *BV* to *PP* is considerable less in magnitude. In other words, a million TL increase/decrease in *BV* generates only 0.02 million TL increase/decrease in *PP*. On the other hand, the results reveal that insurance premium productions is chiefly driven by perceived brand trust where a grade increase/decrease in *BT* motivates *PP* to rise/fall by 5.32 million TL.

To increase *BT* might not be too easy as said. Because the brand itself is an intangible asset that firms want to build in customers perception, thoughts, and hearts (Kapferer, 2007), and to create a valuable trust onto that *"virtual asset"* might require a greater experience and a well-established strategy which both are subjected to a longer time. Surely, a *"greater experience"* might be obtained with a greater customer services, and thus customer satisfaction which eventually contributes to brand image and brand loyalty. As well as, a *"well-established strategy"* incorporates these values with higher brand awareness and recognition. At the end, it creates a positive perception in customers thoughts and triggers their willingness to purchase the products of this brand.

Moreover, a strong momentum documented in this study might be addressed to findings of Pappu et al. (2005) who argue that new customers tend to prefer the firms that already have a large number of customers in portfolio, and Bilgili et al. (2008) who observe that insurance customers in Turkey are too sticky with one company, so that they have very low willingness to shift to another company. This, I believe, itself creates a solid momentum.

## 5. Conclusion

The paper examines how brand dynamics affect insurance premium productions in Turkey. Using a panel of 31 cross-sections over 2005-2015, it documents that the chief driving force behind premium productions is perceived trust of customers onto brands. It reveals that a unit (grade) increase in brand trust augments premium productions of firms by 5.32 million TL. Besides, the analysis also shows that brand value of firms is statistically significant determined of premium production, but its impact size is relatively low than brand trust. Speaking numerically, a million TL increase in brand value of a firm generates only 0.02 million TL additional premium outputs.

On the other hand, the study finds a strong momentum caused by past years premium production which means that 85% of previous years premium production continues to appear in current premium production total. This might be addressed to a higher loyalty-stickiness in Turkey as Bilgili et al. (2008) state, as well as to Pappu et al. (2005) who asserts that a greater customer size of a company itself helps to attract new customers due to *"bandwagon effect"* which creates a momentum. Because even though majority of customers consider themselves acting-rationally, they prefer to act collectively, following the crowds in the market irrationally being victimized by their own emotions of hope, fear, and ambiguity.

# Appendix-A

**Table 5.** *Numerical Transformation of Letter Grade Ratings*

| Brand-Finance Brand Trust Rating | Rating Description | My Numeric Transformation |
|---|---|---|
| AAA+ | *Prime* | 100.00 |
| AAA | *Prime* | 97.50 |
| AAA- | *Prime* | 95.00 |
| AA+ | *High grade* | 92.50 |
| AA | *High grade* | 90.00 |
| AA- | *High grade* | 87.50 |
| A+ | *Upper Medium grade* | 85.00 |
| A | *Upper Medium grade* | 82.50 |
| A- | *Upper Medium grade* | 80.00 |
| BBB+ | *Lower Medium grade* | 75.00 |
| BBB | *Lower Medium grade* | 72.50 |
| BBB- | *Lower Medium grade* | 70.00 |
| BB+ | *Speculative* | 67.50 |
| BB | *Speculative* | 65.00 |
| BB- | *Speculative* | 62.50 |
| B+ | *Highly Speculative* | 60.00 |
| B | *Highly Speculative* | 57.50 |
| B- | *Highly Speculative* | 55.00 |
| CCC+ | *Substantial Risks* | 50.00 |
| CCC | *Substantial Risks* | 47.50 |
| CCC- | *Substantial Risks* | 45.00 |
| CC+ | *Extremely Speculative* | 42.50 |
| CC | *Extremely Speculative* | 40.00 |
| CC- | *Extremely Speculative* | 37.50 |
| C+ | *Default Imminent* | 35.00 |
| C | *Default Imminent* | 32.50 |
| C- | *Default Imminent* | 30.00 |
| D | *In Default* | 25.00 |

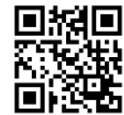